\documentclass[aps,jcp,preprint,superscriptaddress]{revtex4}
\usepackage[utf8]{inputenc}

\usepackage{amsmath, amssymb}
\usepackage{tabularx,booktabs,array,dcolumn}
\usepackage{setspace}

\usepackage{graphicx,psfrag,subfigure}
\usepackage{color}

\usepackage[all]{xy}

\usepackage{enumitem}
\usepackage{mdframed}
\usepackage{xcolor}
\usepackage{rotating}

\newcommand{\ordo}[1]{\mathcal{O}({#1})}
\def\eps{\varepsilon}
\def\mel{m_\text{e}}
\def\massu{m_\text{u}}
\def\iim{\text{i}}
\def\eem{\text{e}}
\def\htot{\widehat{H}}
\def\hel{H_\text{e}}
\def\ddim{{d}}

\usepackage{color}

\usepackage[english]{babel}
\usepackage{pgf, amsmath, amsfonts, amsthm,amsbsy, mathrsfs, latexsym} 
\usepackage{srcltx, hyperref, stackrel}
  
\newcommand{\epsi}{\varepsilon}
\newcommand{\E}{{\mathrm{e}}}
\newcommand{\I}{\mathrm{i}}
 \newcommand{\R}{ \mathbb{R} }
\newcommand{\C}{ \mathbb{C} }

\newcommand{\Or}{{\mathcal{O}}}

\newcommand{\Hi}{{\mathfrak{H}}}
\newcommand{\Hie}{{\mathfrak{H}_{\rm e}}}
\newcommand{\Em}{\end{pmatrix}}
  \newcommand{\ph}{\varphi}

\newtheorem*{theorem*}{Theorem}

\newtheorem*{lemma*}{Lemma}

\newcommand{\He}{{H_{\rm e}}}
\newcommand{\K}{{\widehat K}}
\renewcommand{\H}{{\widehat H}}
\newcommand{\Q}{{\widehat Q}}
\renewcommand{\S}{{\widehat S}}
\renewcommand{\Re}{{\mathcal{R}}}
\newcommand{\A}{{\widehat A}}
\newcommand{\Pih}{{\widehat \Pi}}
\newcommand{\HBON}{{\H_{P}^{(0)}} } 
\newcommand{\HBOn}{{\H_{P}^{(n)}} }

\newcommand{\HeffN}{{{\widehat{\mathcal{H}}}_{P}^{(0)}} } 
\newcommand{\Heffn}{{{\widehat{\mathcal{H}}}_{P}^{(n)}} } 
\newcommand{\HeffZ}{{{\widehat{\mathcal{H}}}_{P}^{(2)}} } 
\newcommand{\HeffD}{{{\widehat{\mathcal{H}}}_{P}^{(3)}} } 

\newcommand{\Hmzero}{{\widehat{\boldsymbol{H}}}_{P}^{(0)}} 
\newcommand{\Hmtwo}{{\widehat{\boldsymbol{H}}}_{P}^{(2)}} 

\newcommand{\od}{{\rm OD}}
\newcommand{\dia}{{\rm D}}
\renewcommand{\L}{\mathscr{L}}
\newcommand{\Li}{\mathscr{I}}
\newcommand{\x}{R}
\newcommand{\y}{r}

\def\kepsi{\tfrac{1}{\epsi}\widehat{K}}
\def\Ht{\widehat{\mathcal{H}}}
\def\Htp{\widehat{h}}

\newcommand{\X}{{\widehat X}}
\newcommand{\Y}{{\widehat Y}}

\begin{document}

\title{%
Effective non-adiabatic Hamiltonians for the quantum nuclear motion
over coupled electronic states \\
}
\author{Edit M\'atyus}
\email{matyus@chem.elte.hu}
\affiliation{Institute of Chemistry, ELTE, E\"otv\"os Lor\'and University, P\'azm\'any P\'eter s\'et\'any 1/A, H-1117 Budapest, Hungary}
\author{Stefan Teufel}
\email{stefan.teufel@uni-tuebingen.de}
\affiliation{Fachbereich Mathematik, Universit\"at T\"ubingen, Auf der Morgenstelle 10, 72076 Tübingen, Germany}
%

\begin{abstract}
\noindent %
The quantum mechanical motion of the atomic nuclei is considered 
over a single- or a multi-dimensional subspace of electronic states 
which is separated by a gap from the rest of the electronic spectrum
over the relevant range of nuclear configurations.
The electron-nucleus Hamiltonian is block-diagonalized 
up to $\ordo{\epsi^{n+1}}$ through a unitary transformation 
of the electronic subspace
and the 
corresponding $n$th-order effective Hamiltonian is derived 
for the quantum nuclear motion.
Explicit but general formulae are given for the second- and 
the third-order corrections.
As a special case, the second-order Hamiltonian corresponding 
to an isolated electronic state
is recovered which contains the coordinate-dependent mass-correction terms 
in the nuclear kinetic energy operator.
For a multi-dimensional, explicitly coupled electronic band,
the second-order Hamiltonian contains the 
usual BO terms and non-adiabatic corrections 
but generalized mass-correction terms appear as well.
These, earlier neglected terms, perturbatively account 
for the outlying (discrete and continuous) electronic states not included in
the explicitly coupled electronic subspace.
\end{abstract}

\maketitle

%
%
\clearpage
\section{Introduction}
\noindent
Molecules are central paradigms of chemistry. 
They acquire unique features as physical objects
due to the three orders of magnitude difference in the mass of their constituent 
particles, the electrons and the atomic nuclei.

Significant improvements in the energy resolution
of spectroscopy experiments \cite{BeHoAgDeScMe18,ChHuJu18}, 
developed or adapted for the molecular domain, 
provide us with new pieces of information which can be deciphered, if
a similarly precise and accurate theoretical description 
becomes available. 
For this purpose, 
it is necessary to re-consider the usual approximations used in quantum chemistry,
in particular, the Born--Oppenheimer (BO)
and the non-relativistic approximations.
These two approximations give rise to `effects' which play
a role at the presently available experimental energy resolution. Further `effects'
may also be visible, for example, 
the interaction between the molecule and the quantized photon field
are more and more appreciated as significant corrections to the molecular energy 
at this resolution \cite{PiJe09}. 
Molecules have a large number of sharp spectral transitions, which can be measured experimentally to high precision.
The interplay of the many small (or often not so small) effects (may) show up 
differently for the transitions between the different dynamical domains, 
so we cannot rely on the cancellation of the small effects, 
but their explicit computation \cite{PiJe09,PuKoPa17,WaYa18}, and hence further development
of molecular quantum theory is necessary.

As to the coupling of the quantum mechanical motion of the electrons and the atomic nuclei
benchmark energies and wave functions can be obtained by the explicit, 
variational solution of the 
few-particle Schrödinger equation \cite{SuVaBook98,CaBuAd03,BuLeStAd09cpl,MaRe12,Ma13,chemrev13,rmp13,PaKo18,MuMaRe18,MuMaRe18b,Ma19rev}. 
We call this direction pre-Born--Oppenheimer (pre-BO) theory, because
it completely avoids the BO separation, nor does it evoke 
the concept of a potential energy surface. Obviously, a pre-BO computation captures `all'
non-adiabatic `effects'.
Although all bound and low-lying resonance states of 
the three-particle H$_2^+=\lbrace \text{p}^+,\text{p}^+,\text{e}^- \rbrace$
molecular ion have been recently reported to an outstanding precision \cite{Ko18},
already for four- and five-particle systems \cite{MuMaRe18b}
the explicit many-particle solution 
is typically limited to a few selected states due to the increased computational cost
and other methodological challenges.

In order to compute (reasonably) accurate energies and wave functions over
a broad dynamical range, we look for effective non-adiabatic Hamiltonians. 

There is a vast literature about dominant non-adiabatic features \cite{MeTr82,PaCeKo88,Ya96}, such as 
conical intersections, the geometric phase effect,
or Jahn--Teller systems.
Practical diabatization procedures \cite{PaCeKo88,ViEi04,KaAvGr16} have been developed
which make it possible to couple close-coming electronic states
(to a good approximation) without the explicit knowledge of 
non-adiabatic coupling vectors, 
which are tedious to compute and burdensome to interpolate for larger
systems.
These effects are sometimes called first-order non-adiabatic effects and
represent qualitatively important features for the molecular dynamics.

In the case of an isolated electronic state the dynamics is 
well described using a single potential energy surface. 
In order to obtain more accurate results, one would need to couple
an increasing number of electronic states.
These additional, explicitly coupled states would give small but non-negligible 
contributions to the molecular energy when studied under high resolution. 
Tightly converging the rovibrational (rovibronic) 
energy by increasing the number of explicitly coupled electronic states is
inpractical (or impossible, since one would need to include also continuum electronic
states). 

A correction which is often computed
is the diagonal Born--Oppenheimer correction (DBOC), 
which gives a mass-dependent contribution to the potential energy surface (PES). 
It has been (empirically) observed that in rovibrational
computations carried out on a single potential energy surface, 
it is `better' (in comparison with experiments) 
to use the atomic mass, especially for heavier atoms, instead of the nuclear mass, 
which would have been rigorously dictated by the BO approximation. 
The difference between the atomic and the nuclear mass is small, it
is the mass of the electrons. This empirical adjustment of the mass of the nuclei
used in the rovibrational kinetic energy operator
has been supplemented with the argument that
attaching the electrons mass to the nuclear mass approximately accounts
for small, `secondary' non-adiabatic effects \cite{Ku07}. 
The empirical adjustment is motivated by the picture 
that the electrons `follow' the atomic nuclei in their motion.

For an isolated electronic state, 
the effective rovibrational Hamiltonian including the rigorous mass-correction terms
has been derived and re-derived in a number of independent and different (perturbative) 
procedures \cite{BuMo77,BuMo80,AFHaUm94,Sch01H2p,Sch01H2O,PaKo09,prx17}
over the past decades and were numerically computed 
for a few systems 
\cite{BuMcMo77,AFHaUm94,BaSaOdOg05,HoSzFrRePeTy11,PaKo12,SaJeOg05,PrCeJeSz17,Ma18nonad,Ma18He2p}.
These mass-correction terms perturbatively account for the effect
of all other electronic states on the rovibrational motion.

\begin{figure}
  \includegraphics[scale=1.]{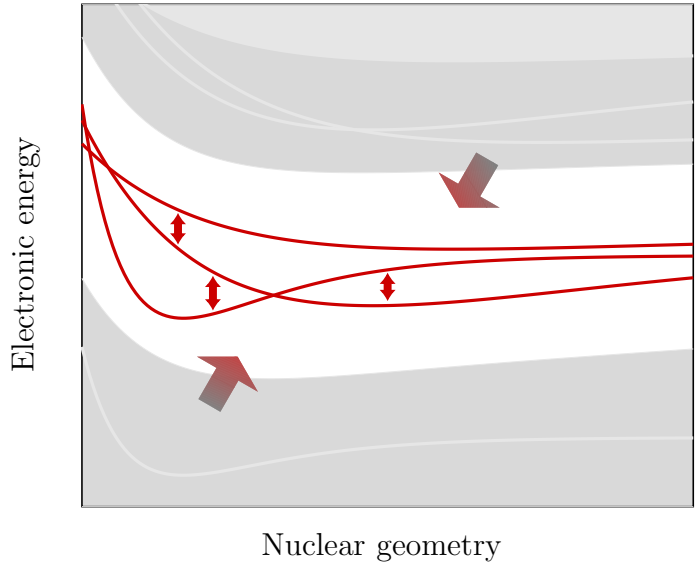}
  \caption{%
  Schematic plot of the electronic energy with respect to the nuclear geometry
  to visualize	 the aim of the present work: formulation of an effective non-adiabatic Hamiltonian for 
  the quantum nuclear motion over an explicitly coupled electronic band,
  which is separated by a gap and decoupled perturbatively 
  from the outlying (discrete or continuous) electronic spectrum.
  }
\end{figure}

We would like to have similar perturbative corrections, 
for a system which is governed by not only a single but 
by a few close-lying electronic states, which 
are explicitly, non-adiabatically coupled, but
which are distant (separated by a finite gap) from 
the rest of the (discrete and continuous) electronic states over the relevant range
of the nuclear coordinates.

To the best of our knowledge, the explicit formulae have never been derived 
for an electronic band which includes multiple electronic states, but 
all the necessary ideas and techniques have been available in the literature,
in particular, in relation with the space-adiabatic theory of 
quantum mechanics \cite{MaSo02,Te03,PaSpTe07, MaSo09, WeLi93}, but 
also other rigorous approaches to compute higher-order corrections to 
the Born-Oppenheimer approximation have been developed, most notably in Ref.~\cite{Ha88}. 
The techniques we use in this article are somewhat reminiscent of van Vleck's
perturbation theory and contact transformation 
often used in chemistry and physics.
We use here a compact and powerful notation which will allow us to obtain 
not only second- but also third-order correction formulae 
for a single or multi-dimensional (non-adiabatically coupled) electronic subspace.

We believe that the explicit formulation of the effective non-adiabatic Hamiltonians for
coupled electronic states, including the earlier missing kinetic (or mass) correction terms, 
will be useful for the chemical physics community. 
Their numerical application assume the computation of 
non-adiabatic coupling vectors. 
Earlier work in which the mass-correction terms were computed for a single
electronic state, \emph{e.g.,} Refs.~\cite{Ma18nonad,Ma18He2p}, can be generalized 
for a multi-state band, so numerical applications will probably follow this theoretical work in the near future.

\vspace{0.5cm}
\paragraph*{Summary of the main result.} 
At the end of this introduction, we summarize the main result of the paper to 
help orientation in the rather technical sections to come. In the following sections, 
we derive the general form of the effective non-adiabatic Hamiltonian $\widehat{\boldsymbol{H}}^{(2)}$ 
for a group of  $d$ electronic levels  $E_1(\x),\ldots,E_d(\x)$ that are separated by a gap  
from the rest of the spectrum. Our analysis implies for example that its eigenvalues approximate 
the eigenvalues of the full molecular Hamiltonian up to order $\epsi^3$, where $\epsi = \sqrt{\frac{m}{M}}$ is 
the square root of the mass ratio of electron and nuclear mass. It thus captures {\em all}   second-order contributions.
It is important to note that the perturbative expansion is carried out without 
assuming small nuclear momenta, so the nuclear kinetic energy $\|\epsi\nabla_R\psi\|^2$ is of order one and 
not of order $\epsi^2$.

After choosing  $d$ electronic  states $\psi_1(\x),\ldots,\psi_d(\x)$
that are smooth functions of $\x$ and pointwise form an orthonormal basis of the selected electronic subspace, 
the projection onto which we denote by $P(R)$, \emph{(i.e.,}\ an adiabatic or diabatic basis set for the selected electronic subspace),
 the effective non-adiabatic Hamiltonian $\widehat{\boldsymbol{H}}^{(2)}$ takes the form of an operator 
 acting on   wave functions  on the nuclear configuration space $\R^{3N}$ that take  values in $\C^d$,  and thus can be written as a 
  $d\times d$-matrix of operators $(\widehat{\boldsymbol{H}}^{(2)})_{\alpha\beta}$ acting on functions on $\R^{3N}$:
\begin{align*}
  (\widehat{\boldsymbol{H}}^{(2)})_{\alpha\beta}
  &=\sum_{i,j=1}^{3N}  
  \left[\tfrac{1}{2}\left( -\I \epsi\partial_i \mathbf{1}+  \epsi\mathbf{A}_i\right) 
  \left(\delta_{ij}\mathbf{1} - \epsi^2 \mathbf{{M}}_{ij}\right)
  \left( -\I   \epsi\partial_j \mathbf{1}+  \epsi\mathbf{A}_j\right) \right]_{\alpha\beta}
  +(\mathbf{E} + \epsi^2\mathbf{\Phi})_{\alpha\beta} \,.
  \end{align*}
Here the boldface objects are $(d\times d)$ matrix-valued functions on the nuclear configuration space, with $(\mathbf{1})_{\alpha\beta}:= \delta_{\alpha\beta}$ denoting the identity matrix, and the others  given as follows in terms of the electronic  states $\psi_1(\x),\ldots,\psi_d(\x)$.

The coefficients of the non-abelian Berry connection, are as expected,
$
\mathbf{A}_{\alpha\beta,i}(\x) =  -\I \langle \psi_\alpha(\x)| \partial_i \psi_\beta(\x) \rangle 
$.
The `diabatic' electronic level matrix becomes 
$\mathbf{E}_{\alpha\beta} (\x) = 
  \langle 
    \psi_\alpha  (\x)| \He(\x) | \psi_\beta(\x)
  \rangle$, 
where $ \He(\x)$ is the electron Hamiltonian for fixed nuclear configuration $\x$.
The second-order diagonal correction is 
$
\mathbf{\Phi}_{\alpha\beta}(\x)  = \tfrac{1}{2} \sum_{i=1}^{3N}   \langle\partial_i \psi_\alpha(\x)| P^\perp(\x)| \partial_i \psi_\beta(\x) \rangle  
$,
where $P^\perp(\x) = 1- P(\x)$ projects on the orthogonal complement of the selected electronic subspace, \emph{i.e.,}\ on the orthogonal complement of the span of  $\psi_1(\x),\ldots,\psi_d(\x)$.

While the matrix versions of terms discussed up to now could have been easily guessed from the single band ($d=1$) case, the determination of the second-order mass correction matrix requires the systematic perturbation approach developed in the following sections. 
The resulting expression is
\[
 \mathbf{M}_{\alpha\beta,ij}  = \sum_{a,b=1}^d \langle \psi_\alpha  | P_a
    (\partial_j P)  (\Re_a + \Re_b) (\partial_i P) P_b
  | \psi_\beta \rangle\,, \]
  where for better readability we dropped the argument $\x$ in all the functions. 
Here $ \Re_a(\x)  := (\He(\x) - E_a(\x))^{-1} \,P^\perp(\x)$ is the reduced resolvent of the level $E_a(\x)$ acting as a bounded operator on the range of $P^\perp(\x)$, and $P_a(\x)$ is the projection onto the eigenspace of $\He(\x)$ corresponding to the eigenvalue $E_a(\x)$. In the special case that $\psi_1(\x),\ldots,\psi_d(\x)$ form an adiabatic basis set, i.e.\
$ \He(\x) \psi_\alpha(\x) = E_\alpha(\x) \psi_\alpha(\x)$ for $\alpha=1,\ldots,d$, the expression for the mass correction term   simplifies to $\mathbf{M}_{ab,ij} (\x)  =  \langle\partial_j\psi_a (\x)|
 \Re_a (\x)+ \Re_b(\x)    | \partial_i\psi_b(\x)\rangle$.

\section{Hamiltonian, coupling parameter, and operator orders}
\paragraph{Molecular Hamiltonian.}
The molecular Hamiltonian is the sum of the nuclear kinetic energy acting on the nuclear coordinates $\tilde R$, the electron kinetic energy acting on the electronic coordinates $r$, and the Coulomb interaction terms, (in Hartree atomic units, $\mel=\hbar=1$)
\begin{align}
  \htot 
  &=
  -\sum_{i=1}^{N} \frac{1}{2m_i} \Delta_{\tilde R_i}
  -\sum_{k=1}^{n} \frac{1}{2} \Delta_{r_k}  
  +V(\tilde R,r)\,.
\end{align}

By absorbing the different masses of the atomic nuclei   in
the 
mass-scaled Cartesian coordinates $  R_i=M_i^{-1/2} \tilde R_i$, where
$M_i$ are the nuclear masses in atomic mass units \cite{GreenBook07}, 
one can  re-write the nuclear kinetic energy operator as 
\begin{align}
  -\sum_{i=1}^{N} \frac{1}{2m_i} \Delta_{\tilde R_i}  
  &=
  -\epsi^2 \sum_{i=1}^{N} \frac{1}{2M_i} \Delta_{\tilde R_i} 
  =
  -\frac{\epsi^2}{2} \sum_{i=1}^{N} \Delta_{  R_i} \nonumber \\
  &=
  -\frac{\epsi^2}{2} \Delta 
  =
  -\frac{\epsi^2}{2} (\epsi\nabla)(\epsi\nabla) 
  =
  -\frac{1}{2} \sum_{j=1}^{3N} (\epsi\partial_j)^2 \,. 
  \label{eq:epsiconv}
\end{align}
and we label $\epsi^2=\mel/\massu \ll 1$ the conversion factor between
the electronic and the atomic mass scale, which is characteristic
to the three orders of magnitude mass difference of the electrons and the 
atomic nuclei.  
 
Using the common notation $\hel(R) := -\sum_{k=1}^{n} \frac{1}{2} \Delta_{r_k}  
  +V(  R,r)$  for the electronic Hamiltonian we can re-write the electron-nucleus Hamiltonian
into the compact form
\begin{align}
  \htot 
  =
  -\frac{\eps^2}{2}\Delta + \hel(R) =: \K + \hel(R)\,,
\end{align}
which highlights the $\epsi^2$ scale-separation (and coupling)
between the electron-nucleus quantum mechanical motion.
Also note that capital letters without hat label operators that are functions
of $R$, \emph{i.e.,} they act fibrewise (pointwise in $R$) on the electronic Hilbert space, such as
$\hel$. All other operators (which include differential operators of $R$) 
are labelled with a wide hat, and the  nuclear kinetic energy $\K$ is an example for this type of operators.

\paragraph{Counting operator orders.}
During the course of this work, we will perform an asymptotic expansion of operators in powers of the small parameter $\epsi$. Since many of the operators appearing in the calculations are unbounded, we emphasize that we are interested in the action of operators on typical molecular wave functions with energies of order $\Or(1)$.
In particular,  the nuclear kinetic energy $-\frac{\epsi^2}{2} \Delta = \frac{1}{2}\,\widehat{p}^{\,2}$ and thus also the 
nuclear momentum operator  $\widehat{p}=-\I\epsi\nabla$ are of order $\Or(1)$ (instead of $\Or(\epsi)$, which one could na\"ively think). This is because a typical molecular wave function shows oscillations with respect to the nuclear coordinates on a spatial scale of order $\epsi^{-1}$ and thus it has derivatives of order $\epsi^{-1}$. 
However, when $\widehat{p}$ acts on a smooth, perhaps operator-valued, function $f(R)$ of the nuclear coordinates, 
we have a quantity $\Or(\epsi)$, indeed: 
\begin{align}
[\widehat{p}, f(R) ]\psi 
= [\widehat{p}, f(R)] \psi
= \widehat{p} (f\psi) - f \widehat{p}\psi
= (\widehat{p}f) \psi 
=-\I\epsi (\nabla f)(R) \psi
\end{align}
These observations will be important to remember for the following calculations.

\section{The zeroth-order non-adiabatic Hamiltonian: 
truncation error and a strategy for reducing this error \label{ch:zeroth}}
\noindent Let $\{E_a(\x), a=1,\ldots,\ddim \}$ be a finite set of eigenvalues 
of the electronic Hamiltonian $\He(\x)$
that are isolated by a finite gap~\footnote{%
A precise formulation of the gap assumption is as follows. We say that there is a gap of size $g>0$ 
in the relevant region $\Omega\subset \R^{3N}$ of nuclear configuration space if there are two continuous functions 
$f_\pm:\Omega\to \R$ such that dist$(f_\pm(R),\sigma(\He(R)))\geq g/2$ 
such that the interval $I(R):= [f_-(R),f_+(R)]$ satisfies $I(R)\cap \sigma(\He(R)) = \{E_1(R), \ldots, E_d(R)\}$. 
Then, all correction terms (Berry phase, diagonal correction, effective mass) are asymptotically small as $\epsi\to 0$ 
when choosing any smooth diabatic basis $\{\psi_1(R),\ldots,\psi_d(R)\}$. 
In a real problem, of course, $\epsi$ is fixed by the physical parameters and smallness of 
the correction terms is an indicator for a sufficiently large spectral gap.} 
from the rest of the spectrum 
over the relevant range of nuclear configurations
and denote by $P_a(\x)$ the corresponding spectral projections. 
Then $P(\x) = \sum_{a=1}^\ddim P_a(\x)$ 
projects onto the selected electronic subspace and $P^\perp := 1-P$ is the projection onto 
the orthogonal complement.

The full electron-nucleus Hamiltonian can be written in a block form as
\begin{equation}\label{BO0}
\H 
= 
\begin{pmatrix} 
  P\H P & P \H P^\perp \\ 
  P^\perp \H P & P^\perp \H P^\perp 
\end{pmatrix} , 
\end{equation}
where $\HBON:=P\H P$ corresponds to the `usual' 
non-adiabatic Hamiltonian~\footnote{%
Note that we follow here the physical chemistry terminology in which a description is called non-adiabatic 
if it concerns more than one electronic states which are coupled among each other. 
Compare this terminology with the mathematics' naming conventions, in which BO or `adiabatic' 
is commonly used for a description in which a (single or multi-dimensional) electronic subspace 
is not coupled with the rest of the electronic spectrum.}.
In numerical computations $\HBON$ is usually represented over some adiabatic or diabatic basis, whereas the 
the off-diagonal  $P \H P^\perp$ and $P^\perp \H P$ blocks have been neglected in earlier work.

The following calculation shows that the off-diagonal terms are indeed small
(remember the gap condition at the beginning of this section), more precisely,  of order $\epsi$:
\begin{align}
 P^\perp \H P 
 &=    
 P^\perp (\K+\hel) P \nonumber \\
 &= 
 P^\perp\K P  \quad\quad ([P,\hel]=0) \nonumber \\
 &= 
 P^\perp[\K, P]P  \quad\quad (PP=P \text{ and } P^\perp P=0)\nonumber\\
 &= 
 -\tfrac{\epsi^2}{2} P^\perp [\Delta,P] P\nonumber\\
 &=
 -\tfrac{1}{2} P^\perp (\epsi^2\Delta P - P \epsi^2\Delta) P \nonumber \\
 &= 
 -\tfrac{1}{2} P^\perp (%
   \epsi\nabla\cdot \epsi(\nabla P) + \epsi(\nabla P)\cdot \epsi\nabla
   ) P \nonumber  \\
 &= 
 -\tfrac{\epsi}{2}  P^\perp \left(%
   \epsi\nabla\cdot P' + P'\cdot \epsi\nabla 
  \right)P \nonumber \\
 &=:
 -\epsi P^\perp \widehat P'_\nabla P \; ,
 \label{eq:dirder}
\end{align}
where for any fibred operator $A$ we use the abbreviations $A':=\nabla A$, and 
the symmetrized directional derivative of $A$ is
\begin{align}
  \widehat  A'_\nabla := \tfrac{1}{2}(\epsi\nabla\cdot A' + A'\cdot \epsi\nabla) = \tfrac{\I}{2}(\widehat{p} \cdot A' + A'\cdot \widehat{p})\,,
  \label{eq:symdirder}
\end{align}
\emph{i.e.,} $A'$ is again a fibred operator, while $\widehat A'_\nabla$ is a first-order differential operator in the nuclear coordinates of $\Or(1)$.
For the upper-right off-diagonal block we find similarly
\begin{align}
  P \H P^\perp = \epsi P \widehat P'_\nabla P^\perp\,. 
\end{align}
Hence, the off-diagonal part of the nuclear kinetic energy operator $\widehat K$ is of order $\epsi$ and we introduce the abbreviation
\begin{eqnarray}  
\label{QDef}
\Q 
&:=&  \tfrac{1}{\epsi}\left(P \K P^\perp+P^\perp \K P\right) \nonumber \\
&=&
P\widehat P'_\nabla   P^\perp  -     P^\perp \widehat P'_\nabla   P \, ,
\end{eqnarray}
which, as explained above, 
is an operator of $\Or(1)$.
Using
$\Q$ and the properties of $P$ and $P^\perp$, the off-diagonal blocks of the Hamiltonian
are written in the compact form
\begin{align}
  P^\perp \H P = \epsi P^\perp \Q P
  \quad \text{and} \quad
  P\H P^\perp = \epsi P \Q P^\perp \, ,
\end{align}
which are both of $\Or(\epsi)$, and thus
\begin{equation}\label{BO0}
\H 
= 
\begin{pmatrix} 
  \HBON  & \epsi P \Q P^\perp \\[1mm]  \epsi P^\perp \Q P  & P^\perp \H P^\perp 
\end{pmatrix}
=
\begin{pmatrix} \HBON  & 0 \\ 0  & P^\perp \H P^\perp \end{pmatrix} + \Or(\epsi) \,.
\end{equation}
This expression confirms the well-known fact, 
the off-diagonal non-adiabatic couplings are small 
(for a group of bands separated by a gap
from the rest of the electronic spectrum). 
As a consequence, 
the spectrum of $\HBON$ provides an $\Or(\epsi)$ approximation, at least locally in energy, to the spectrum of $\H$.  
More precisely, within a neighbourhood of order $\epsi$ around any  
spectral value of $\HBON$, there is also a spectral value of the full Hamiltonian~$\H$~\footnote{One simple way to see this is as follows: Let $\psi$ be a normalized eigenfunction of $\HBON$, i.e.\ $(\HBON - E)\psi =0$ and $\|\psi\|= \|P\psi\| = 1$. Then   $\chi:= (\H -E)\psi$ has norm $\|\chi\| = \epsi \|P^\perp \Q \psi\|$  of order $\epsi$ and thus $\|(\H -E)^{-1} \frac{\chi}{\|\chi\|}\| =\frac{\|\psi\|}{\|\chi\|} \gtrsim \frac{1}{\epsi}$. Hence dist$(E, {\rm spec}(\H)) = \frac{1}{\| (\H -E)^{-1}\|}\lesssim \epsi$. For spectral values in the continuous spectrum of $\HBON$ one can use the same argument with a Weyl sequence instead of an eigenfunction.}.

To obtain a better approximation, we replace the projection $P$ by a slightly `tilted' projection $\Pih$, such that the off-diagonal terms in the  block-decomposition of $\H$ with respect to $\Pih$ are lower order, namely
\begin{align}
  \H 
  = 
  \begin{pmatrix} 
    \Pih  \H \Pih  & \Pih  \H \Pih ^\perp \\ 
    \Pih ^\perp \H \Pih  & \Pih ^\perp \H \Pih ^\perp 
  \end{pmatrix}
  =
  \begin{pmatrix} 
    \HBOn  &  0 \\ 
    0  & \Pih ^\perp \H \Pih^\perp 
    \end{pmatrix}
    +\Or(\epsi^{n+1})
  \label{eq:blockdiagham}
\end{align}
for some $n\geq 1$.
The projection $\Pih$ is obtained from $P$ through 
a near-identity unitary transformation,
\begin{align}
  \Pih = \E^{\I \epsi \S} P \E^{-\I \epsi \S}\,,
\end{align}
where the generator $\S \approx \A_1 + \eps \A_2 + \eps^2 \A_3 + \ldots$
will be determined exactly by the condition that the off-diagonal elements in 
Eq.~(\ref{eq:blockdiagham}) are of order $\epsi^{n+1}$.
The $n$th-order effective  Hamiltonian $\HBOn$  obtained in this way will give 
an  $\Or(\epsi^{n+1})$ approximation to the spectrum of $\H$. 
The physical picture behind the projection $\Pih$  is the following. The range of the adiabatic projection $P$ is spanned by states of the form $\Psi(R,r) = \ph(R) \psi_a(R,r)$, where $\psi_a(R,r)$ are eigenstates of the electronic Hamiltonian $\hel(R)$ for the 
clamped nuclear configuration $\x$,  $\hel(R)\psi_a(R,\cdot)= E_a(R)\psi_a(R,\cdot)$. However, since the nuclei are also moving, the molecular eigenstates are only approximately but not exactly  of this local product form. 
Loosely speaking, the state of the electrons depends also on the momenta of the nuclei.
This effect is taken care of by 
slightly tilting the projection $P$ into the projection $\Pih$.

For computing eigenvalues of the $n$th-order effective Hamiltonian $\HBOn$,
it will be more appropriate to consider the  unitarily equivalent (and thus isospectral) operator:
\begin{align}\label{eq:heffdef}
  \Heffn 
  :=  
  \E^{- \I \epsi \S}\HBOn \E^{\I \epsi \S} \nonumber \\
  =  
  \E^{- \I \epsi \S}\Pih  \H \Pih \E^{\I \epsi \S} \nonumber \\
  = 
  P \E^{- \I \epsi \S}  \H \E^{\I \epsi \S} P,
\end{align}
which, by choosing a basis representation for $P$, 
will provide an $n$th-order effective 
Hamiltonian for the quantum nuclear motion corresponding
to the selected electronic subspace.
The eigenvectors of $\HBOn$ and $\Heffn$ are related 
by the unitary transformation $\E^{\I \epsi \S}$.

%
%
\section{Calculation of the $n$th-order transformation matrix}
\noindent 
We would like to achieve an $\Or(\epsi^{n+1})$ block diagonalization
of the electron-nucleus Hamiltonian by conjugating it with an appropriate 
unitary operator $\eem^{\iim \eps \S}$, 
a strategy  that is somewhat reminiscent of what is known in 
chemical physics as van Vleck's perturbation theory and contact transformation.
The actual procedure we use is known as adiabatic perturbation theory 
in the mathematical literature \cite{MaSo02,Te03,PaSpTe07, MaSo09}, but was applied before in physics also 
in the context of the Born--Oppenheimer approximation in Ref.~\cite{WeLi93}. 
While the quoted references are, with the exception of Ref.~\cite{PaSpTe07}, use pseudodifferential calculus 
and are therefore quite technical and demanding, our approach is rather elementary and uses  
a general and compressed notation. As a consequence, we are able to derive
with relatively little effort explicit expressions not only for 
$\HeffZ$ but also for $\HeffD$ and 
not only for a single but for a finite number of coupled electronic states.
Note that $\HeffZ$ for a single electronic state was explicitly calculated earlier, \emph{e.g.,} 
in Ref.~\cite{WeLi93,Te03,PaSpTe07,BuMo77,PaKo09,prx17},
although in some cases without carefully paying attention to the subtleties of counting 
operator orders in relation with the nuclear momenta (see Sec. II.b).

The simplicity and generality of our derivation 
is based on the concept and simple algebraic properties of diagonal and off-diagonal operators
and an explicit expression for the  inverse  of the quantum Liouvillian \emph{(vide infra)} acting on operators which do not contain nuclear differential operators, as to be explained in the upcoming section.

\subsection{Technical preliminaries}
\noindent %
Before constructing $\S$ and $\Heffn$, we introduce the concept
of (off)diagonal operators and the (inverse) Liouvillian as well as some of their properties,
which will become useful during the course of the calculations. Further useful relationships 
are collected in the Appendix.

\subsubsection{Diagonal and off-diagonal operators}
We define
the diagonal ($\dia$) and the off-diagonal ($\od$) 
parts of a linear operator $\widehat A$ with respect to the orthogonal projections $P$ and $P^\perp = 1-P$ as
\begin{align}
\widehat A^\dia := P\widehat AP + P^\perp \widehat AP^\perp
\quad\mbox{and}\quad 
\widehat A^\od := P\widehat AP^\perp + P^\perp \widehat AP\, ,
\end{align}
respectively.
From this definition, a number of simple relationships follow immediately. For example, it holds for all operators $\A$ and $\widehat B$ that
\begin{align}
  [\widehat A^\dia ,P] =[\widehat A^\dia ,P^\perp] = 0
  \quad\mbox{and}\quad 
  [\widehat A^\od ,P] = P^\perp \widehat A P - P\widehat AP^\perp \,,
\end{align}
and
\begin{align}\label{eq:algebra}
[\A^\dia,\widehat B^\dia]^\od = 0 \,,\quad [\A^\dia,\widehat B^\od]^\dia = 0\,, \quad \mbox{and} \quad [\A^\od,\widehat B^\od]^\od = 0\,.
\end{align}
Finally,  note that $\Q$ defined in Eq.~(\ref{QDef}) is
related to the off-diagonal part of the nuclear kinetic energy as
\begin{align}
  \K^\od = 
  \epsi \Q. 
  \label{eq:keoff}
\end{align}

\subsubsection{Commutators, the quantum Liouvillian and its inverse}
Given an operator $\widehat B$, let us define the linear mapping  
\begin{align}
\widehat A\mapsto \L_{\widehat B}(\widehat A) := -\I [\widehat B,\widehat A]\,,
\end{align}
whenever the commutator is well defined (for the operators we consider in the following sections, 
no problems with operator domains occur). 
For $\widehat B = \He$ we call this mapping
the quantum Liouvillian of $\widehat{A}$:
\begin{align}
  \L_\He(\widehat{A}) = -\I [\He, \widehat{A}].
  \label{eq:liou}
\end{align}
Since $\He$ commutes with $P$ and $P^\perp$,  the Liouvillian $\L_\He$ preserves (off)diagonality, \emph{i.e.,}\ 
\begin{equation}\label{eq:LiouvilleDia}
\L_\He(\widehat A^\dia) = \L_\He(\widehat A)^\dia
\quad\mbox{and}\quad 
\L_\He(\widehat A^\od) = \L_\He(\widehat A)^\od\,.
\end{equation}
Furthermore, $\L_\He$ is invertible on the space of  fibred off-diagonal operators (labelled without any hat).
Explicitly, for $B:=B^\od$   its inverse is
\begin{align}
\Li_\He(B)=\I\sum_{a=1}^d \left( \Re_a   B P_a - P_a B   \Re_a
\right)\, , 
  \label{eq:linv}
\end{align}
where 
\begin{align}
  \Re_a(\x) := (\He(\x) - E_a(\x))^{-1} P^\perp(\x) 
  \label{eq:redres}
\end{align}
is the reduced resolvent (with the nuclear coordinate dependence shown explicitly). 
Note that, due to the gap condition, $ (\He(\x) - E_a(\x))^{-1}$ is indeed a bounded operator when restricted to states in the orthogonal complement of the range of $P(R)$. 

In the following lines, we check that Eq.~(\ref{eq:linv}) is indeed the inverse Liouvillian on off-diagonal fibred operators, but
we carry out this calculation for a general operator, $\widehat B$, which 
may contain also the nuclear momentum operator (highlighted with a hat in the notation):
\begin{eqnarray}
\L_\He(\Li_\He (\widehat B)) 
&=& 
-\I  \left[\He,  \I\sum_a \left( \Re_a   \widehat B P_a - P_a \widehat B   \Re_a \right)\right] \nonumber \\
&=& 
\sum_a \left[\He, \left( \Re_a   \widehat B P_a - P_a \widehat B   \Re_a \right)\right] \nonumber \\
&=& 
\sum_a \left[\He- E_a, \left( \Re_a   \widehat B P_a - P_a \widehat B   \Re_a \right)\right]    
+ 
\sum_a \left( \Re_a   [  E_a,\widehat B] P_a - P_a [  E_a,\widehat B]  \Re_a \right) \nonumber \\
&=&  
\sum_a \left( P^\perp \widehat B P_a + P_a \widehat B P^\perp\right)
+ 
\sum_a \left( \Re_a   [  E_a,\widehat B] P_a - P_a [  E_a,\widehat B]  \Re_a \right) \nonumber \\
& =&  
\widehat B^\od+ \sum_a \left( \Re_a   [  E_a,\widehat B] P_a - P_a [  E_a,\widehat B]  \Re_a\right)\, ,
\label{eq:invcalc}
\end{eqnarray}
where $E_a(R)$ is the electronic energy.
For off-diagonal, fibred operators $\Li_\He$ is indeed the (exact)
inverse of $\L_\He$, since $ [E_a,B]=0$. Otherwise,
it is an approximation to the inverse with an error depending on 
the value of the commutator $[E_a,\widehat B]$ .
For example, if $\widehat B=\K^\od$, the commutator in Eq.~(\ref{eq:invcalc}) is 
$[E_a(R),\widehat K^\od]=\epsi[E_a(R),\Q]$, and thus
$\L_{\hel}(\Li_{\hel}(\K^\od))= \K^\od + \Or(\epsi)$, so the inverse is obtained with 
an $\Or(\epsi)$ error.

%
%
\subsection{Conditions for making the off-diagonal block of the Hamiltonian lower order}
\noindent As explained in the previous section, in order to reduce the off-diagonal coupling between the selected  electronic subspace $P$
and its  orthogonal complement $P^\perp$, 
we will choose the self-adjoint  operator $\S=\A_1+\epsi\A_2+\ldots$
such that the off-diagonal part (coupling) within the $\Pih$-block decomposition of the  Hamiltonian $\H$, 
Eq.~(\ref{eq:blockdiagham}), is small:
\begin{align}
  \Pih^\perp\H\Pih \stackrel{!}{=} \Or(\epsi^{n+1}) \; .
  \label{eq:smalloff}
\end{align}
Due to hermiticity, this condition also implies that $\Pih\H\Pih^\perp$ is of $\Or(\epsi^{n+1})$.
To construct explicitly the operators $\A_1,\A_2,\ldots$,
it is more practical to use the unitary transform 
\begin{align}
\E^{-\I\epsi \S} \Pih^\perp \H \Pih  \E^{\I\epsi \S}
&= 
P^\perp \E^{-\I\epsi \S} \H \E^{\I\epsi \S} P \nonumber \\
&= 
P^\perp \Ht P \nonumber \\
&\stackrel{!}{=}\Or(\epsi^{n+1})
\label{eq:condition2}
\end{align}
of condition \eqref{eq:smalloff},
where the transformed Hamiltonian is defined as
\begin{align}
  \Ht = \E^{-\I\epsi \S} \H \E^{\I\epsi \S} \; .
\end{align}
In the language of diagonal and off-diagonal operators, 
the condition of Eq.~(\ref{eq:condition2}) is fulfilled 
if the \emph{off-diagonal part} of the transformed Hamiltonian, 
$\Ht^\od=P^\perp \Ht P + P \Ht P^\perp$, is \emph{small:}
\begin{align}
  \Ht^\od \stackrel{!}{=}\Or(\epsi^{n+1})\, ,
  \label{eq:Htod}
\end{align}
which will be our working equation to determine the operators $\A_1, \ldots, \A_n$  up to 
the $(n+1)$st order in $\epsi$.
Then, using these operators, an explicit 
expression will be derived for the relevant block of the transformed Hamiltonian, namely of 
$
\Heffn=P\,\Ht P
$  as defined in Eq.~(\ref{eq:heffdef}) and explained in Section~\ref{ch:zeroth}.

%
%
\subsection{Reduction of the off-diagonal coupling: determination of $\A_1,\ldots,\A_n$ \label{ch:htod}}
The transformed electron-nucleus Hamiltonian is expanded in terms of 
increasing powers of $\epsi$ (see Appendix) as
\begin{align}
  \Ht
  &=
  \E^{-\I\epsi\S} 
  \H
  \E^{\I\epsi\S} \nonumber \\
  &=
  \H 
  + \epsi \L_\S(\H)
  + \tfrac{\epsi^2}{2} \L_\S(\L_\S(\H))
  + \tfrac{\epsi^3}{6} \L_\S(\L_\S(\L_\S(\H)))
  + \ldots
\end{align}
For the electronic Hamiltonian the  expansion up to $\Or(\epsi^4)$ is
\begin{align}
  \E^{-\I\epsi\S}
  \hel
  \E^{\I\epsi\S}
  = & \;\hel  + \epsi \L_{\A_1}(\hel)  + 
    \epsi^2 \left( %
      \L_{\A_2}(\hel) + \tfrac{1}{2} \L_{\A_1}(\L_{\A_1}(\hel))
    \right)
    \nonumber \\
  &\;+ 
    \epsi^3 \left( %
      \L_{\A_3}(\hel) 
      + \tfrac{1}{2} \L_{\A_1}(\L_{\A_2}(\hel)) \right. \nonumber \\
  &\qquad\;\left.
      + \tfrac{1}{2} \L_{\A_2}(\L_{\A_1}(\hel))
      + \tfrac{1}{6} \L_{\A_1}(\L_{\A_1}(\L_{\A_1}(\hel)))
    \right)
    +\Or(\epsi^4) \; .
\end{align}
Anticipating that commutators of the form $[\A_j, \K]$ are of order $\epsi$, 
 the expansion for the nuclear kinetic energy term up to the same order is
\begin{align}
  \E^{-\I\epsi\S}
  \K
  \E^{\I\epsi\S}     = &\;  
 \K   + \epsi^2 \L_{\A_1}(\kepsi) \nonumber \\
  &\;+ \epsi^3 \left(%
    \L_{\A_2}(\kepsi) + \tfrac{1}{2}\L_{\A_1}(\L_{\A_1}(\kepsi)) 
  \right)
  + \Or(\epsi^4) \; .
\end{align}
Note that $\K=\K^\dia+\epsi\hat{Q}$, so its diagonal part is of leading order, 
while the  off-diagonal  part is $\Or(\epsi)$.

Thus, the transformed Hamiltonian has an asymptotic expansion in powers of $\epsi$,
\begin{align}
  \Ht &= 
  \Htp_0 + \epsi \Htp_1 + \epsi^2 \Htp_2 +   \ldots +  \epsi^n \Htp_n + \Or(\epsi^{n+1})\,,
\end{align}
with
\begin{align}
  \Htp_0 &= \K + \hel\,,
  \label{eq:htzero}
\end{align}
\begin{align}
  \Htp_1 &= \L_{\A_1}(\hel)  = -\L_{\hel}(\A_1)  \,,
  \label{eq:htone}
\end{align}
\begin{align}
  \Htp_2 &= \L_{\A_2}(\hel) + \tfrac{1}{2} \L_{\A_1}(\L_{\A_1}(\hel)) + \L_{\A_1}(\kepsi) \nonumber \\
        &= -\L_{\hel}(\A_2) - \tfrac{1}{2} \L_{\A_1}(\L_{\hel}(\A_1)) + \L_{\A_1}(\kepsi)\,,
  \label{eq:httwo}        
\end{align}
and
\begin{align}
  \Htp_3 &= \L_{\A_3}(\hel) 
           + \tfrac{1}{2} \L_{\A_1}(\L_{\A_2}(\hel))
           + \tfrac{1}{2} \L_{\A_2}(\L_{\A_1}(\hel))
           + \tfrac{1}{6} \L_{\A_1}(\L_{\A_1}(\L_{\A_1}(\hel))) \nonumber \\
         &\qquad\quad+ \L_{\A_2}(\kepsi)
           + \tfrac{1}{2} \L_{\A_1}(\L_{\A_1}(\kepsi)) \nonumber \\
        &= -\L_{\hel}(\A_3) 
           - \tfrac{1}{2} \L_{\A_1}(\L_{\hel}(\A_2))
           - \tfrac{1}{2} \L_{\A_2}(\L_{\hel}(\A_1))
           - \tfrac{1}{6} \L_{\A_1}(\L_{\A_1}(\L_{\hel}(\A_1))) \nonumber \\
         &\qquad\quad+ \L_{\A_2}(\kepsi)
           + \tfrac{1}{2} \L_{\A_1}(\L_{\A_1}(\kepsi)) \,.
  \label{eq:htthree}           
\end{align}

To reduce the off-diagonal coupling, we will now proceed by induction.
Assuming that 
$\A_1,\A_2,\ldots,\A_{n-1}$ have been chosen such that 
\begin{align}
(\Ht^{(n-1)} )^\od := \left( \sum_{i=0}^{n-1}\epsi^i\Htp_{i}\right) ^\od=  \sum_{i=0}^{n-1}\epsi^i\Htp_{i}  ^\od=:\epsi^n \widehat B_{n-1}
\end{align}
 is  $\Or(\epsi^n)$, we will fix $\A_n$ such that $(\Ht^{(n)} )^\od= \sum_{i=0}^{n}\epsi^i\Htp_{i}^\od=:\epsi^{n+1} \widehat B_{n}$ is  $\Or(\epsi^{n+1})$.

We mention already at this point that fulfillment of this sequence 
of requirements will fix only  the  off-diagonal part $\A_i^\od$ of each $\A_i$. 
The  diagonal parts, $\A_i^\dia$, generate merely rotations within the subspaces $P$ and $P^\perp$,
but do not affect the (de)coupling. Hence,   we set $\A_i^\dia=0$ ($i=1,2,\ldots$).
With this choice, $\A_1,\A_2,\ldots,\A_n$ are completely determined
by the requirement, Eq.~\eqref{eq:Htod}.

\vspace{0.5cm}
\noindent\textbf{Zeroth-order off-diagonal (OD) terms:} 
The off-diagonal part of ${\Htp}_0$, Eq.~(\ref{eq:htzero}), 
\begin{align}
  \Htp^\od_0 = \K^\od = \epsi \Q =:\epsi \widehat B_0
\end{align}
is of order $\epsi$, hence decoupling is automatically fulfilled at this order. 

\vspace{0.5cm}
\noindent\textbf{First-order OD terms:} 
In the next step we require
\begin{align}
\Htp_0^\od + \epsi \Htp_1^\od  \;&=\; \epsi( \widehat B_0 + \Htp_1^\od) 
 \;=\; \epsi ( \Q -\L_{\hel}(\A_1)^\od)  \nonumber \\
  &= \;
 \epsi ( \Q -\L_{\hel}(\A_1^\od)) \;\stackrel{!}{=}\; \epsi^2 \widehat B_1  ,\label{eq:laonedef}
\end{align}
where we used Eq.~\eqref{eq:LiouvilleDia} and the fact that $\A_1=\A_1^\od$.
This condition can be fulfilled by choosing 
\begin{align}
  \A_1: = \Li_{\hel}(\Q).
  \label{eq:defAone}
\end{align}
Inserting Eq.~(\ref{eq:defAone}) back into Eq.~(\ref{eq:laonedef}) and 
using Eq.~(\ref{eq:invcalc}), we find that 
\begin{align}
  \epsi \widehat B_1
&=
 \Q - \L_{\hel}(\Li_{\hel}(\Q)) \nonumber \\
&= -
  \sum_a \left( \Re_a   [  E_a,\widehat Q] P_a - P_a [  E_a,\widehat Q]  \Re_a\right), \nonumber \\
&=  -
  \epsi \sum_a  \left( \Re_a   \left[  E_a,\tfrac{1}{\epsi}\Q \right] P_a - P_a \left[  E_a,\tfrac{1}{\epsi}\Q \right]  \Re_a\right) \nonumber \\
&= -
  \epsi \sum_a   E_a'\cdot  \left( \Re_a  P'   P_a +P_a  P' \Re_a\right) \; .
  \label{eq:defBone}
\end{align}
This explicit expression for $\widehat B_1$ will be required for 
calculating the third-order terms in the effective Hamiltonian $\HeffD$
(see Eqs.~(\ref{eq:exAtwo})--(\ref{eq:corrthree})).

\vspace{0.5cm}
\noindent\textbf{Second-order OD terms:}
We require
\begin{align}
\Htp_0^\od + \epsi \Htp_1^\od + \epsi^2 \Htp_2^\od
   &=  \epsi^2 \widehat B_1 + \epsi^2 \Htp_2^\od\nonumber \\
   &= \epsi^2\left( \widehat  B_1
   -\L_{\hel}(\A_2)^\od 
   - \tfrac{1}{2} \L_{\A_1}(\L_{\hel}(\A_1))^\od 
   + \L_{\A_1}(\kepsi)^\od 
   \right) \nonumber \\
   &=\epsi^2\left( \widehat  B_1
   -\L_{\hel}(\A_2)
   + \L_{\A_1}(\kepsi^\dia)
 \right) \;\stackrel{!}{=}\; \epsi^3 \widehat B_2\; ,\label{eq:h2od}
\end{align}
where in the third equality we used the algebraic relations of Eq.~(\ref{eq:algebra}).
Again, we  solve this equation for $\A_2$  
using the approximate inverse Liouvillian, Eq.~(\ref{eq:linv}):
\begin{align}
  \A_2 
  := 
  \Li_{\hel}(%
    \L_{\A_1}(\kepsi^\dia)
    + \widehat B_1
  )\; .
  \label{eq:defAtwo}
\end{align}
The explicit expression for the  remainder $\widehat B_2$ could be determined, if needed,
through the calculation of the $\Or(\epsi)$ error term, Eq.~(\ref{eq:invcalc}),
from the approximate inversion:
\begin{align}
  \epsi \widehat B_2  
  &=\widehat  B_1
  -\L_{\hel}(\A_2) 
  +
  \L_{\A_1}(\kepsi^\dia)  \nonumber \\  
  &= \widehat  B_1
  -\L_{\hel}(
    \Li_{\hel}(
    \L_{\A_1}(\kepsi^\dia)
    +\widehat  B_1)
  ) 
  +
  \L_{\A_1}(\kepsi^\dia) \,.
\end{align}

\vspace{0.5cm}
\noindent\textbf{Third-order OD terms:}
By the same reasoning we require
\begin{align}
\widehat  B_2 + \Htp^\od_3     &= \widehat B_2
        -\L_{\hel}(\A_3)^\od
           - \tfrac{1}{2} \L_{\A_1}(\L_{\hel}(\A_2))^\od
           - \tfrac{1}{2} \L_{\A_2}(\L_{\hel}(\A_1))^\od
           \nonumber \\
         &\qquad - \tfrac{1}{6} \L_{\A_1}(\L_{\A_1}(\L_{\hel}(\A_1)))^\od + \L_{\A_2}(\kepsi)^\od
           + \tfrac{1}{2} \L_{\A_1}(\L_{\A_1}(\kepsi))^\od
          \nonumber \\
   &=\widehat B_2 
        -\L_{\hel}(\A_3)
           - \tfrac{1}{6} \L_{\A_1}(\L_{\A_1}(\L_{\hel}(\A_1))) 
         + \L_{\A_2}(\kepsi^\dia)
           + \tfrac{1}{2} \L_{\A_1}(\L_{\A_1}(\Q))\nonumber\\
           & \stackrel{!}{=} \epsi \widehat B_3
            \;,\label{eq:h3od}
\end{align}
where we used that the last two terms in the first  line are zero (off-diagonal part of diagonal operators).
Again, we make the left hand side of Eq.~(\ref{eq:h3od}) small by solving the equation for $\A_3$  
using the approximate inverse Liouvillian, Eq.~(\ref{eq:linv}):
\begin{align}
  \A_3
  :=
  \Li_{\hel}\left(%
  \widehat B_2
    - \tfrac{1}{6} \L_{\A_1}(\L_{\A_1}(\L_{\hel}(\A_1))) 
    + \L_{\A_2}(\kepsi^\dia)
    + \tfrac{1}{2} \L_{\A_1}(\L_{\A_1}(\Q))
  \right)
\end{align}
and the  remainder term $\widehat B_3$ can be determined by the direct calculation of the 
$\Or(\epsi)$ error of the inversion using Eq.~(\ref{eq:invcalc}).

It is obvious how to continue this induction to arbitrary orders. However, as we are only interested in explicit expressions for the effective Hamiltonians up to third order, we refrain from stating the general induction explicitly.

%
%
\subsection{Second- and third-order Hamiltonians \label{ch:htdia}}
In this section we calculate the leading terms in the expansion of the $n$th-order effective Hamiltonian 
\begin{align}
  \Heffn
  &=
  \sum_{j=0}^n
  \epsi^j P\,\Htp_j P + \Or(\epsi^{n+1})
\end{align}
up to and including $P\,\Htp_3 P$, and thus obtain explicit expressions for 
the second and the third-order effective  Hamiltonians, $\HeffZ$ and $\HeffD$.
To this end, we first calculate the diagonal parts $\Htp_j^\dia$ and then, in a second step, project onto the range of $P$.

\vspace{0.5cm}
\noindent\textbf{Zeroth-order diagonal (D) terms:}
\begin{align}
  \Htp_0^\dia 
  &= \K^\dia+
  \hel^\dia = \K^\dia+ \hel 
  \label{eq:heffzero}
\end{align}

\vspace{0.5cm}
\noindent\textbf{First-order D terms:} Recalling that $\A_j= \A_j^\od$ for all $j\geq 1$ and 
the algebraic relations in Eq.~(\ref{eq:algebra}), we find, in particular, that
\begin{align}
  \Htp_1^\dia 
  &=  
  -\L_{\hel}(\A_1)^\dia= -\L_{\hel}(\A_1^\dia)= 0 \,.
  \label{eq:heffone}
\end{align}

\vspace{0.5cm}
\noindent\textbf{Second-order D terms:} Similarly, since $\widehat B_j=\widehat B_j^\od$  for all $j\geq 1$, we find
\begin{align}
  \Htp_2^\dia 
  &\;=\; 
  -\L_{\hel}(\A_2)^\dia - \tfrac{1}{2} \L_{\A_1}(\L_{\hel}(\A_1))^\dia + \L_{\A_1}(\kepsi)^\dia \nonumber \\
  &\;=\; 
  - \tfrac{1}{2} \L_{\A_1}(\L_{\hel}(\A_1)) + \L_{\A_1}(\Q) \nonumber \\
  &\;=\;   
  - \tfrac{1}{2} \L_{\A_1}(\Q - \epsi \widehat B_1) + \L_{\A_1}(\Q)  \nonumber \\
  &\;=\;   
  \tfrac{1}{2} \L_{\A_1}(\Q) +\tfrac{\epsi }{2} \L_{\A_1}(\widehat B_1)
  \label{eq:hefftwo}        
\end{align}

\vspace{0.5cm}
\noindent\textbf{Third-order D terms:}
\begin{align}
  \Htp_3^\dia 
        &= -\,\L_{\hel}(\A_3)^\dia 
           - \tfrac{1}{2} \L_{\A_1}(\L_{\hel}(\A_2))^\dia
           - \tfrac{1}{2} \L_{\A_2}(\L_{\hel}(\A_1))^\dia
           - \tfrac{1}{6} \L_{\A_1}(\L_{\A_1}(\L_{\hel}(\A_1)))^\dia \nonumber \\
         & \quad \,+ \L_{\A_2}(\kepsi)^\dia
           + \tfrac{1}{2} \L_{\A_1}(\L_{\A_1}(\kepsi))^\dia \nonumber \\
        &= 
           - \tfrac{1}{2} \L_{\A_1}(\L_{\hel}(\A_2))
           - \tfrac{1}{2} \L_{\A_2}(\L_{\hel}(\A_1))
           + \L_{\A_2}(\Q)
           + \tfrac{1}{2} \L_{\A_1}(\L_{\A_1}(\kepsi^\dia)) \nonumber \\
        &= 
           - \tfrac{1}{2} \L_{\A_1}(\L_{\A_1}(\kepsi^\dia)+\widehat B_1-\epsi\widehat  B_2)
           - \tfrac{1}{2} \L_{\A_2}(\Q-\epsi\widehat  B_1)
           + \L_{\A_2}(\Q)
           + \tfrac{1}{2} \L_{\A_1}(\L_{\A_1}(\kepsi^\dia)) \nonumber \\
        &= -
           \tfrac{1}{2} \L_{\A_1}(\widehat B_1)
           +\tfrac{1}{2} \L_{\A_2}(\Q)
           +\Or(\epsi)        \,.
\label{eq:heffhree}           
\end{align}
By combining these expressions, we obtain the second- and 
the third-order effective Hamiltonians as
\begin{align}\label{eq:HeffZ}
  \HeffZ
  &:=
 P\K P + P\hel P +  \tfrac{\epsi^2}{2} P\L_{\A_1}(\Q) P  + \Or(\epsi^3) \; 
\end{align}
and
\begin{align}\label{eq:HeffD}
  \HeffD
  &:=
 P\K P + P\hel P +  \tfrac{\epsi^2}{2} P\L_{\A_1}(\Q) P + \tfrac{\epsi^3}{2} P\L_{\A_2}(\Q) P  
  + \Or(\epsi^4) \; ,
\end{align}
respectively.
We note that in the third-order correction, 
the 
$- \tfrac{1}{2} \L_{\A_1}(\widehat B_1)$ remainder from 
second order,
cancels the 
$\tfrac{1}{2} \L_{\A_1}(\widehat B_1)$ 
third-order term.
In the following subsection, we continue with inserting 
the explicit formulae for $\A_1$, $\A_2$, and $\Q$ into the compact expressions of $\HeffZ$ and $\HeffD$ just obtained.

%
%
\subsection{More explicit expressions for the second- and 
the third-order non-adiabatic Hamiltonian corrections}
%
\paragraph{Second-order correction.}
Using the explicit expression for the inverse Liouvillian, Eq.~(\ref{eq:linv}),
Eq.~(\ref{eq:defAone}) yields
\begin{align}
\A_1 =  \I \sum_{a=1}^d \left( \Re_a   \Q  P_a - P_a \Q    \Re_a\right) \, .
\end{align}
Thus, the $\Or(\epsi^2)$ correction term of the effective Hamiltonian in Eq.~(\ref{eq:HeffZ}) is
\begin{eqnarray}\nonumber
  \tfrac{\epsi^2}{2} P   \L_{\A_1}(\Q ) P 
  &=& 
  -\tfrac{\epsi^2}{2} \sum_{a,b=1}^d \left( P_b\Q \Re_a   \Q  P_a + P_a \Q    \Re_a \Q P_b \right) \\ 
  &=&   
  \tfrac{\epsi^2}{2} \sum_{a,b=1}^d \left( P_b P'_\nabla \Re_a  P'_\nabla  P_a + P_a P'_\nabla    \Re_a P'_\nabla P_b
\right)\, ,
\label{SecOrdEx}
\end{eqnarray}
where we inserted Eq.~(\ref{QDef}) for $\Q$ and 
used the fact that the reduced resolvent $\Re_a$ acts only in the $P^\perp$ subspace. 
(Recall that $P'_\nabla$ is defined by Eq.~(\ref{eq:symdirder}).)
Since the commutator of $\epsi\nabla$ in $P'_\nabla$ with smooth, fibred operators 
yields higher-order terms in $\epsi$, we can further simplify $\HeffZ$ to 
\begin{align}
  \HeffZ 
&=
  P\K P 
  + P\He P \nonumber \\
&\qquad  + \tfrac{\epsi^2}{2} %
 \sum_{j,i=1}^{3N}  
  \sum_{a,b=1}^d  \Big(%
  P_b (\epsi\partial_j) (\partial_j P)  \Re_a    (\partial_i P)  (\epsi\partial_i)P_a
  +P_a (\epsi\partial_j) (\partial_j P)  \Re_a    (\partial_i P)  (\epsi\partial_i)P_b   
\Big) \nonumber \\
&=
  P\K P 
  + P\He P   \nonumber \\
&\qquad
  +\tfrac{\epsi^2}{2}  %
  \sum_{j,i=1}^{3N}  
  \sum_{a,b=1}^d
  (\epsi\partial_j) \Big(%
    P_b (\partial_j P)  \Re_a  (\partial_i P)  P_a 
   +P_a (\partial_j P)  \Re_a  (\partial_i P)  P_b 
  \Big) (\epsi\partial_i) +\Or(\epsi^3) \nonumber \\
&=
  P\K P 
  + P\He P   
  +\epsi^2  %
  \sum_{j,i=1}^{3N}  
  \sum_{a,b=1}^d 
  (\epsi\partial_j) 
  P_a (\partial_j P)  \tfrac{\Re_a + \Re_b}{2} (\partial_i P)  P_b 
  (\epsi\partial_i) +\Or(\epsi^3)\; .
\label{eq:heffsecfinal}
\end{align}
Note, however, that the $\Or(\epsi^3)$ term does contribute to the third-order effective Hamiltonian 
and can not be neglected when computing  $\HeffD$.

\vspace{1cm}
\paragraph{Third-order correction.}
To obtain an explicit expression for the third-order correction, we need to derive an
explicit expression for 
  $\A_2$, Eq.~(\ref{eq:defAtwo}): 
\begin{align}
  \A_2 
    &=   \Li_\He ( \L_{\A_1}(\tfrac{1}{\epsi}\K^\dia )) +  \Li_\He (\widehat B_1) 
  \; ,
  \label{eq:exAtwo}
\end{align}
which assumes the explicit knowledge of the  first-order remainder term $\widehat B_1$, Eq.~(\ref{eq:defBone}), too.
The first term in Eq.~(\ref{eq:exAtwo}) includes
\begin{eqnarray}
 \L_{\A_1}(\tfrac{1}{\epsi}\K^\dia ) 
 &=& 
 -\I [\A_1,\tfrac{1}{\epsi}\K^\dia]  \nonumber \\
 &=& 
 -\I \left[ \I \sum_{a=1}^d \left( \Re_a   \Q  P_a - P_a \Q  \Re_a\right),\tfrac{1}{\epsi}\K^\dia \right] \nonumber \\
 &=&  
 \tfrac{1}{\epsi}\sum_{a=1}^d \left( \Re_a   \Q  P_a \K P - P^\perp \K   \Re_a    \Q  P_a 
- P_a  \Q    \Re_a \K P^\perp +P\K P_a  \Q     \Re_a \right) 
\end{eqnarray}
and its inverse Liouvillian, Eq.~(\ref{eq:linv}), is 
\begin{eqnarray}
  \Li_\He(\L_{\A_1}(\tfrac{1}{\epsi}\K^\dia )  ) 
  &=&  
  \I \sum_{a,b=1}^d \left(%
    \Re_b  \, \L_{\A_1}(\tfrac{1}{\epsi}\K^\dia )   P_b - P_b  \,\L_{\A_1}(\tfrac{1}{\epsi}\K^\dia )  \Re_b
  \right) \nonumber \\
  &=&  
  \tfrac{\I}{\epsi} \sum_{b=1}^d \left( \Re_b (\Re_a    \Q   P_a \K - \K   \Re_a   \Q   P_a )  P_b - P_b (\K P_a  \Q     \Re_a - P_a  \Q     \Re_a \K ) \Re_b \right)
  \nonumber \\
  &=& 
  \tfrac{\I}{\epsi}\sum_{a=1}^d ( P_a  \Q    \Re_a \K \Re_a-\Re_a   \K   \Re_a   \Q  P_a  ) \nonumber \\
&& +\;  \tfrac{\I}{\epsi}\sum_{a,b=1}^d ( \Re_b\Re_a \Q  P_a\K P_b- P_b  \K P_a \Q     \Re_a \Re_b  ) .
\label{eq:Atwoterm1}
\end{eqnarray}
The inverse Liouvillian of $\widehat B_1$, Eq.~(\ref{eq:defBone}), is 
\begin{eqnarray}
  \Li_\He(\widehat B_1)  
  &=&  
  \I\sum_{a=1}^d \left( \Re_a  \widehat  B_1 P_a - P_a \widehat B_1   \Re_a \right) \nonumber \\
  &=&  
 - \I\sum_{a,b=1}^d    E_b'\cdot \left( \Re_a   \Re_b    P' P_b P_a - P_a P_b   P' \Re_b   \Re_a \right) \nonumber \\
  &=&  
 - \I\sum_{a=1}^d   E_a'\cdot \left( \Re_a   \Re_a    P' P_a - P_a    P' \Re_a   \Re_a \right) \; .
\label{eq:Atwoterm2}  
\end{eqnarray}
Next, the explicit expression for $\A_2$, obtained as the sum of Eqs.~(\ref{eq:Atwoterm1}) and (\ref{eq:Atwoterm2}),
is used to expand the third-order correction as
\begin{eqnarray}
\nonumber
\tfrac{1}{2}P\L_{\A_2}(\Q)P 
&=& 
-\tfrac{\I}{2 } P[\A_2,  \Q  ]P \\
\nonumber 
&=&  \tfrac{1}{2\epsi}\sum_{a=1}^d ( P \Q  \Re_a   \K   \Re_a   \Q   P_a + P_a  \Q    \Re_a \K \Re_a \Q  P )\\
\nonumber 
&& -\;  \tfrac{1}{2\epsi}\sum_{a,b=1}^d ( P_b  \K P_a  \Q     \Re_a \Re_b \Q  P  +P \Q  \Re_b\Re_a \Q  P_a\K P_b )\\
&& +\; \tfrac{1}{2} \sum_{a=1}^d    E_a'\cdot \left( P \Q  \Re_a   \Re_a    P' P_a + P_a   P' \Re_a   \Re_a  \Q  P \right) \; .
\label{thirdorder3}
\end{eqnarray}
By working out the second sum of Eq.~(\ref{thirdorder3}), we obtain
\begin{eqnarray}
\lefteqn{%
\sum_{a,b=1}^d ( P_b  \K P_a  \Q     \Re_a \Re_b \Q  P  +P \Q  \Re_b\Re_a \Q  P_a\K P_b )} \nonumber \\
 &=&  \sum_{a,b=1}^d ( P_b  [\K, P_a]  \Q     \Re_a \Re_b \Q  P  +P \Q  \Re_b\Re_a \Q  [P_a, \K] P_b ) \nonumber \\
 && + \sum_{a=1}^d ( P_a  \K    \Q     \Re_a \Re_a \Q  P  +P \Q  \Re_a\Re_a \Q   \K P_a ) \nonumber \\
 &=&  \sum_{a,b=1}^d ( P_b  [\K, P_a]  \Q     \Re_a \Re_b \Q  P  +P \Q  \Re_b\Re_a \Q  [P_a, \K] P_b ) \nonumber \\
 && + \sum_{a=1}^d( P_a  [\K,    \Q]     \Re_a \Re_a \Q  P  +P \Q  \Re_a\Re_a [\Q,   \K] P_a ) \nonumber \\
  && + \sum_{a=1}^d ( P_a      \Q \K     \Re_a \Re_a \Q  P  +P \Q  \Re_a\Re_a \K \Q  P_a ) \nonumber \\
   &=&  \sum_{a,b=1}^d ( P_b  [\K, P_a]  \Q     \Re_a \Re_b \Q  P  +P \Q  \Re_b\Re_a \Q  [P_a, \K] P_b ) \nonumber \\
 && +\sum_{a=1}^d ( P_a  [\K,    \Q]     \Re_a \Re_a \Q  P  +P \Q  \Re_a\Re_a [\Q,   \K] P_a ) \nonumber \\
  && + \sum_{a=1}^d ( P_a      \Q [\K ,    \Re_a] \Re_a \Q  P  +P \Q  \Re_a[\Re_a, \K] \Q  P_a ) \nonumber \\
  && + \sum_{a=1}^d ( P_a      \Q    \Re_a\K \Re_a \Q  P  +P \Q  \Re_a\K \Re_a\Q  P_a ) \; ,
\end{eqnarray}
where the last expression exactly cancels the first term in Eq.~(\ref{thirdorder3}), and thus the correction term
at third order is
\begin{eqnarray}
\tfrac{1}{2}P\L_{\A_2}(\Q)P 
&=&  
-\tfrac{1}{2 }\sum_{a,b=1}^d \left( P_b  [\tfrac{1}{\epsi}\K, P_a]  \Q     \Re_a \Re_b \Q  P  +P \Q  \Re_b\Re_a \Q  [P_a, \tfrac{1}{\epsi}\K] P_b \right) \nonumber \\
&& - \tfrac{1}{2 }\sum_{a=1}^d \left( P_a  [\tfrac{1}{\epsi}\K,    \Q]     \Re_a \Re_a \Q  P  +P \Q  \Re_a\Re_a [\Q,   \tfrac{1}{\epsi}\K] P_a \right) \nonumber \\
  && - \tfrac{1}{2 } \sum_{a=1}^d \left( P_a      \Q [\tfrac{1}{\epsi}\K ,    \Re_a] \Re_a \Q  P  +P \Q  \Re_a[\Re_a, \tfrac{1}{\epsi}\K] \Q  P_a \right) \nonumber \\
&& +\; \tfrac{1}{2} \sum_{a=1}^d    E_a'\cdot \left( P \Q  \Re_a   \Re_a    P' P_a + P_a   P' \Re_a   \Re_a  \Q  P \right) \; .
\label{eq:corrthree1}
\end{eqnarray}
The commutators can be evaluated as
\begin{align}
 [\tfrac{1}{\epsi}\K, P_a] = -\tfrac{\epsi}{2} [\Delta, P_a] = -\tfrac{1}{2}(\epsi\nabla\cdot P_a' + P_a' \cdot \epsi\nabla) = -   P_{a\nabla}'\,,
\end{align}
and
\begin{align}
P [\tfrac{1}{\epsi}\K, \Q] P^\perp = -\tfrac{\epsi}{2} P [\Delta,    P  P'_\nabla   P^\perp  -     P^\perp   P'_\nabla    P  ]   P^\perp
   =  - P P''_{\nabla^2} P^\perp    +\Or(\epsi)\,,
\end{align}
with $P''_{\nabla^2} := \epsi  \sum_{i,j=1}^{3N}\partial_j (\partial_j\partial_iP) \epsi \partial_i$, and 
\begin{align}
 P^\perp [\tfrac{1}{\epsi}\K, \Re_a] P^\perp =   -   P^\perp \Re'_{a\nabla}P^\perp\,.
\end{align}
Inserting these   identities into Eq.~(\ref{eq:corrthree1}), we finally obtain for the third-order correction as 
\begin{eqnarray}
\tfrac{\epsi^3}{2}P\L_{\A_2}(\Q)P 
&=&    
- \tfrac{\epsi^3}{2 }\sum_{a,b=1}^d \left( P_b P_{a\nabla}'  P'_\nabla     \Re_a \Re_b P'_\nabla  P  -P P'_\nabla  \Re_b\Re_a P'_\nabla  P_{a\nabla}' P_b \right) \nonumber \\
 && - \tfrac{\epsi^3}{2 }\sum_{a=1}^d \left(  P_a P''_{\nabla^2}    \Re_a \Re_a P'_\nabla  P  -P P'_\nabla  \Re_a\Re_a P''_{\nabla^2}P_a \right)   \nonumber \\
  && - \tfrac{\epsi^3}{2 } \sum_{a=1}^d \left( P_a      P'_\nabla P^\perp\Re'_{a\nabla} \Re_a P'_\nabla  P  -P P'_\nabla  \Re_a\Re'_{a\nabla} P^\perp P'_\nabla  P_a \right) \nonumber \\
&& +\; \tfrac{\epsi^3}{2} \sum_{a=1}^d    E_a'\cdot \left( P  P'_\nabla  \Re_a   \Re_a    P' P_a - P_a   P' \Re_a   \Re_a   P'_\nabla  P \right)
\label{eq:corrthree} \;+\Or(\epsi^4)\; .
\end{eqnarray}
Note that the first three lines are third order in the nuclear momentum, $\widehat p$, 
and the last line is linear in $\widehat p$.

When looking at the second- and the third-order corrections in Eq.~(\ref{eq:heffsecfinal}) and (\ref{eq:corrthree}), one might worry about
singular expressions. Indeed, some of the summands become singular near points of the nuclear configuration space at which eigenvalues within the  set $\{E_a\,|\,a=1,\ldots,d\}$ cross. At these points the single spectral projections $P_a$ might not be differentiable. 
However, as it can be seen from the original expressions, Eqs.~\eqref{eq:HeffZ} and \eqref{eq:HeffD}, for $\HeffZ$ and $\HeffD$,  
the complete expression (the full sum) remains bounded because the singularities in the different summands cancel each other. 
This property might require additional care in numerical computations.

%
%
\section{Basis representation and effective nuclear Hamiltonians}
\noindent
By choosing an electronic (\emph{e.g.,} adiabatic or diabatic) basis set 
$\{ \psi_\alpha, \alpha=1,\ldots,\ddim \}$ for $P\Hie$, 
one can represent a molecular wave function $\Psi$  in the range of $P$, $\Psi\in P\Hi$, as $\Psi(\x,\y) = \sum_{\alpha=1}^{\ddim} \ph_\alpha(\x) \psi_\alpha(\x,\y)$
(where $\Hie$ and $\Hi$ denote the electronic and the molecular Hilbert space, respectively). 
It is common practice to represent the zeroth-order effective Hamiltonian $\HeffN= \HBON = P\H P$ as a matrix operator $\Hmzero$ with respect to such a basis set, which then acts only on the nuclear functions $(\ph_1(\x),\ldots , \ph_{\ddim}(\x))$. This yields, in particular, also the Berry phase and the diagonal BO correction terms (see below). 

In what follows, we construct 
the matrix representation also
for the  second-order Hamiltonian $\HeffZ$.
As a special case, the known mass-correction terms for 
a single, isolated electronic state will be recovered.
The basis representation for the  third-order correction,
Eq.~(\ref{eq:corrthree}), can be worked out along the same lines.
 
\subsection{Basis representation for the second-order, multi-state Hamiltonian\label{ch:basishamtwo}}
\noindent Let us choose  an electronic (\emph{e.g.,} adiabatic or diabatic) basis set 
 $\psi_1(\x),\ldots,\psi_d(\x)$ such that  the $\psi_\alpha$
 are smooth functions of $\x$ and pointwise form an orthonormal basis of the range of $P(\x)$, \emph{i.e.,}
\begin{align}
\langle \psi_\alpha(\x) | \psi_\beta(\x) \rangle=\delta_{\alpha,\beta}\quad \mbox{and}\quad 
 P(\x)=\sum_{ \alpha=1}^\ddim | \psi_\alpha(\x) \rangle \langle \psi_\alpha(\x)  |\,.
\end{align}
Because of the gap condition, such a smooth diabatic basis set always exist \footnote{Strictly speaking, such a smooth diabatic basis set exists over each contractible subset $\Omega\subseteq \R^{3N}$ of the nuclear configuration space $\R^{3N}$ on which the gap condition is satisfied.  This is because the gap condition implies that the projections $P(\x)$ define a smooth rank-$d$ vector bundle over $\Omega$, and a vector bundle over a contractible set always allows for a global trivialisation, that is  for a diabatic basis set as above.}.
However, due to possible crossings within the set of eigenvalues $E_1,\ldots, E_d$, 
it might  {\bf not} be possible to choose $\psi_1(\x),\ldots,\psi_d(\x)$ as smooth functions of $\x$, and 
at the same time, as pointwise eigenfunctions of $\hel(\x)$, \emph{i.e.,} 
\begin{align}\label{adiabaticbasisset}
  \He(\x) \psi_\alpha(\x) = E_\alpha(\x) \psi_\alpha(\x), \quad  \alpha=1,\ldots,\ddim\,\qquad\mbox{is not assumed in general!}
\end{align}
Then, the matrix representation of $\HeffZ$, Eq.~(\ref{eq:heffsecfinal}), 
over $\psi_\alpha$, $\alpha=1,\ldots,\ddim$, results in 
a  matrix  operator $\Hmtwo$ for the quantum nuclear motion with matrix elements 
\begin{align}
  (\Hmtwo)_{\alpha\beta}
  &=
  \langle 
    \psi_\alpha  | 
    \HeffZ 
    |\psi_\beta
  \rangle
 =  
 \langle 
    \psi_\alpha  | \K | \psi_\beta
  \rangle
  +
  \langle 
    \psi_\alpha  | \He | \psi_\beta
  \rangle
  \nonumber \\
&\qquad\qquad 
  +\tfrac{\epsi^2}{2}  %
  \sum_{j,i} \sum_{a,b}
  (\epsi\partial_j) 
  \langle \psi_\alpha  | P_a
    (\partial_j P)  (\Re_a + \Re_b) (\partial_i P) P_b
  | \psi_\beta \rangle
  (\epsi\partial_i) +\Or(\epsi^3)\,.
\end{align}
For the kinetic-energy part, we find
\begin{align}
  \langle 
    \psi_\alpha  | \K | \psi_\beta
  \rangle= -
  \sum_{i}
  \left(
  \tfrac{1}{2}
  (\epsi\partial_i)^2 \delta_{\alpha\beta}
   +\epsi 
    \langle \psi_\alpha| \partial_i \psi_\beta \rangle  
    (\epsi\partial_i)
    + \tfrac{\epsi^2}{2}  \langle \psi_\alpha| \partial_i^2 \psi_\beta \rangle  
    \right)\,.
\end{align}
By introducing the coefficient of the non-abelian Berry-connection
\begin{align}
\mathbf{A}_{\alpha\beta,i}:=  -\I \langle \psi_\alpha| \partial_i \psi_\beta \rangle = \overline{\mathbf{A}_{\beta\alpha,i} }\,,
\end{align}
we find
\begin{align}
\left[\tfrac{1}{2}\left( -\I \epsi \partial_i \mathbf{1}+ \epsi \mathbf{A}_i\right)^2\right]_{\alpha\beta} &=
\left[ -\tfrac{1}{2}(\epsi\partial_i)^2 \mathbf{1}-\I \epsi \mathbf{A}_i(\epsi\partial_i) -
\I\tfrac{\epsi^2}{2} (\partial_i \mathbf{A}_i) + \tfrac{\epsi^2}{2} (\mathbf{A}_i)^2\right]_{\alpha\beta}
\nonumber \\
&= -\tfrac{1}{2}(\epsi\partial_i)^2\delta_{\alpha\beta}
-
\epsi 
    \langle \psi_\alpha| \partial_i \psi_\beta \rangle  
    (\epsi\partial_i) -\tfrac{\epsi^2}{2}  \langle \psi_\alpha| \partial_i^2 \psi_\beta \rangle  \nonumber \\
    &\quad -\tfrac{\epsi^2}{2}  \langle\partial_i \psi_\alpha| \partial_i \psi_\beta \rangle  
    -\tfrac{\epsi^2}{2} \sum_\gamma \langle \psi_\alpha| \partial_i \psi_\gamma \rangle\langle \psi_\gamma| \partial_i \psi_\beta \rangle\,.
\end{align}
With $\langle \psi_\alpha| \partial_i \psi_\gamma \rangle= - \langle\partial_i \psi_\alpha|  \psi_\gamma \rangle$, the last term becomes
\begin{align}
 \sum_\gamma \langle \psi_\alpha| \partial_i \psi_\gamma \rangle\langle \psi_\gamma| \partial_i \psi_\beta \rangle= -  \sum_\gamma  \langle\partial_i \psi_\alpha|  \psi_\gamma \rangle\langle \psi_\gamma| \partial_i \psi_\beta \rangle
 = -  \langle\partial_i \psi_\alpha| P|\partial_i \psi_\beta \rangle  \;,
\end{align}
and thus the kinetic-energy term can be written in the form
\begin{align}
 \langle \psi_\alpha  | \K | \psi_\beta  \rangle
  =\sum_{i=1}^{3N}  
  \left[\tfrac{1}{2}\left( -\I \epsi \partial_i \mathbf{1}+ \epsi \mathbf{A}_i\right)^2\right]_{\alpha\beta}
  +\epsi^2\mathbf{\Phi}_{\alpha\beta}  \,  
\end{align}
with the matrix-valued Berry-connection coefficient $\mathbf{A}_i$ and the matrix-valued  potential energy correction
\begin{align}
\mathbf{\Phi}_{\alpha\beta}(\x) := \tfrac{1}{2} \sum_{i=1}^{3N}   \langle\partial_i \psi_\alpha(\x)| P^\perp(\x)| \partial_i \psi_\beta(\x) \rangle \,.
\end{align}
This latter quantity can be understood as an $\Or(\epsi^2)$ correction to the `diabatic' electronic level matrix
\begin{align}
\mathbf{E}_{\alpha\beta} (\x):= 
  \langle 
    \psi_\alpha  (\x)| \He(\x) | \psi_\beta(\x)
  \rangle \,.
\end{align}
Note that $\mathbf{E}_{\alpha\beta}$ is a diagonal matrix if and only if all $\psi_\alpha$ are eigenvectors of $\He$.
Defining the second-order mass-correction term as
\begin{align}
 \mathbf{M}_{\alpha\beta,ij} := \sum_{a,b=1}^d \langle \psi_\alpha  | P_a
    (\partial_j P)  (\Re_a + \Re_b) (\partial_i P) P_b
  | \psi_\beta \rangle\,,
    \label{eq:hamtwoadbas}
\end{align}
the matrix representation of $\HeffZ$ 
over $\psi_\alpha$, $\alpha=1,\ldots,\ddim$, can be compactly written as
\begin{align}
  (\Hmtwo)_{\alpha\beta}
  &=
  \sum_{i,j=1}^{3N}  
    \left[
      \tfrac{1}{2}\left( -\I \epsi \partial_i \mathbf{1}+ \epsi \mathbf{A}_i\right) 
      \left(\delta_{ij}\mathbf{1} - \epsi^2 \mathbf{{M}}_{ij}\right)
      \left( -\I \epsi \partial_j\mathbf{1}+ \epsi \mathbf{A}_j\right) 
    \right]_{\alpha\beta}
  +(\mathbf{E} + \epsi^2\mathbf{\Phi})_{\alpha\beta} \;+\Or(\epsi^3)\,.
  \label{eq:basHeffZ}
\end{align}
We note that this is the complete second-order non-adiabatic Hamiltonian operator 
for the nuclear motion. It is important to remember
the peculiarities of counting operator orders (Sec. II.c),
which follow from not making the assumption that the nuclear momenta 
(when $-\iim\epsi\partial_R$ acts on the nuclear wave function) are small.
Also note that we used Cartesian coordinates scaled 
with the nuclear mass, Eq.~(\ref{eq:epsiconv}), 
so the derived expressions can be used also for heteronuclear systems 
by making this scaling factor explicit in the numerical computations.

In the special case where $\psi_1,\ldots,\psi_d$ form an adiabatic basis set, 
\emph{i.e.,}\ satisfy Eq.~(\ref{adiabaticbasisset}), 
the expression for $\mathbf{M}_{\alpha\beta,ij}$ simplifies to 
\begin{align*}
\mathbf{M}_{ab,ij} & =  \langle\partial_j\psi_a |
 \Re_a + \Re_b    | \partial_i\psi_b\rangle   \,.  
\end{align*}
To see this, one uses
  that the reduced resolvent, Eq.~(\ref{eq:redres}), contains a projection $P^\perp$, 
and
\begin{align}
  (\partial_j P) P^\perp
  &=
  \sum_{\gamma=1}^\ddim (\partial_j |\psi_\gamma\rangle \langle \psi_\gamma |) P^\perp
  =
  \sum_{\gamma=1}^\ddim |\psi_\gamma\rangle \langle \partial_j \psi_\gamma | P^\perp  \; 
\end{align}
and similarly for its adjoint, 
$P^\perp (\partial_i P) = \sum_{\gamma=1}^\ddim P^\perp |\partial_i \psi_\gamma\rangle \langle \psi_\gamma |$.

In all our expressions the nuclear differential operators 
are written in terms of Cartesian coordinates. The operators can be
transformed to curvilinear coordinates, necessary for efficient rovibrational computations,
similarly to the transformation of the single-state
non-adiabatic Hamiltonian as it was carried out in Ref.~\cite{Ma18nonad} 
using the Jacobi and the metric tensors of the new coordinates.

For the special case of a single electronic state ($\psi_1,E_1$), we are free to choose a real-valued, 
normalized electronic wave function $\psi_1$. Then, the effective operator of the atomic nuclei,
Eq.~(\ref{eq:basHeffZ}), simplifies to
\begin{align}
  (\Hmtwo)_{1,1}
  &= 
  \sum_{i,j=1}^{3N}  
      \tfrac{1}{2}\left( -\I \epsi \partial_i  \right)
      \left(\delta_{ij} - \epsi^2 {M}_{11,ij}\right)
      \left( -\I \epsi \partial_j \right) 
   +E_1 + \epsi^2\Phi_1 +\Or(\epsi^3) \,.
\end{align}
So, we assume that $\psi_1$ is chosen such that $A_1 = -\I \langle\psi_1|\nabla\psi_1\rangle =0$ and 
find for the mass correction term that
\begin{align}
  {M}_{11,ij}
  &=
  2\langle 
    \partial_j \psi_1  
    |  \Re_1 |
    \partial_i \psi_1
  \rangle  \nonumber \\
  &=
  2\langle 
    \partial_j \psi_1  
    |  (\He - E_1)^{-1}(1-P_1) |
    \partial_i \psi_1
  \rangle    
  \; .
\end{align}
This mass-correction function is identical with that used in Ref.~\cite{Ma18nonad}, 
and thus, for a single, isolated electronic state, the known expression of the second-order 
non-adiabatic Hamiltonian is recovered.

%
%
\clearpage
\section{Summary and conclusions}
\noindent %
Molecular wave functions are often approximated on the  subspace $P\Hi$
of the full electron-nucleus $\Hi$ Hilbert space, where
$P$ is the electronic subspace which governs 
the motion of the atomic nuclei.

We have shown that a complete neglect of the complementary electronic subspace $(1-P)\Hi$
introduces an $\Or(\epsi)$ error in the Hamiltonian and also 
in the molecular spectrum ($\epsi$
is the square root of the electron-to-nucleus mass ratio).
We improved upon this $\Or(\epsi)$ approximation, by 
using a near-identity unitary transform of $P$,
$\Pih = \E^{\I \epsi \S} P \E^{-\I \epsi \S}$.
Terms of the self-adjoint  
transformation operator $\S=\A_1+\epsi\A_2+\epsi^2\A_3+\ldots$ were determined
for increasing orders of $\epsi$ by making the coupling, and hence
the error of the molecular energy, $\epsi$ times smaller at every order.
The resulting transformation operators include the momentum operator $\widehat p$ 
of the atomic nuclei, thereby the transformed electronic space $\Pih$, which makes the coupling lower order, 
depends not only on the  nuclear positions $R$ but also on the  nuclear momenta $\widehat p$.
The transformed electronic states adjusted by $\widehat p$ up to order $\widehat p^{n}$, 
achieve a block-diagonalization of $\H$ up to terms of order $\epsi^{n+1}$.
From the transformed, $\Or(\epsi^{n+1})$ block-diagonal Hamiltonian, 
we obtained effective $n$th-order Hamiltonians for the quantum nuclear motion. 
Explicit expressions were derived up to the third-order corrections for a multi-dimensional
electronic subspace.

In particular, the second-order non-adiabatic Hamiltonian contains correction terms 
quadratic in the nuclear momenta, which may be small near the bottom of the electronic band, 
but for highly excited states they can easily dominate the diagonal correction. 
These kinetic energy correction terms can be identified as a coordinate-dependent
correction to the nuclear mass in the nuclear kinetic energy operator.
These earlier neglected `mass-correction terms'
perturbatively account for the effect of the electronic states
not included in the selected, explicitly coupled electronic band.
For a single electronic state the multi-state expressions simplifies
to the known, second-order Hamiltonian including the mass-correction function.

This perturbative decoupling can be used for isolated (groups of) 
electronic states and we believe that at least the second-order, multi-state
expression will soon gain practical applications in rovibronic and 
quantum scattering computations. 
Examples for potential applications include 
the electronically excited manifold of molecular hydrogen---the first steps 
towards these applications are reported in Ref.~\cite{FeMa19b}---,
the predissociation dynamics of H$_3^+$,  
in which the interaction of the electronic ground and excited states 
is thought to play a role, and also the H+H$_2$ reactive scattering system.

%
%
\section{Appendix}
\paragraph{Contact transform of an operator.}
Transformation of an operator $\Y$ with $\E^{\I\epsi\S}$ can be expanded in terms of 
increasing powers of $\epsi$ as
\begin{align}
  {\widehat{\mathcal{Y}}} 
  =
  \E^{-\I\epsi\S}
  \Y
  \E^{\I\epsi\S}
  =
  \Y 
  + \epsi \L_\S(\Y)
  + \tfrac{\epsi^2}{2} \L_\S(\L_\S(\Y))
  + \tfrac{\epsi^3}{6} \L_\S(\L_\S(\L_\S(\Y)))
  + \ldots
\end{align}

\paragraph{Commutator operations with diagonal and off-diagonal operators.}
\begin{align}
  [\X^\dia,\Y]^\dia = [\X^\dia,\Y^\dia] \quad&\text{and}\quad
  [\X^\dia,\Y]^\od = [\X^\dia,\Y^\od] \\
  [\X^\od,\Y]^\dia = [\X^\od,\Y^\od] \quad&\text{and}\quad
  [\X^\od,\Y]^\od = [\X^\od,\Y^\dia] \; .
\end{align}
For the example of the diagonal $\hel$ and the off-diagonal $\A_i$,
we have collected the following identities (relevant for the calculations in the manuscript):
\begin{align} 
  \L_{\hel}(\A_i)^\dia &= 0 \\
  \L_{\hel}(\A_i)^\od &= \L_{\hel}(\A_i) \\[0.3cm]
  \L_{\A_j}(\L_{\hel}(\A_i))^\dia &= \L_{\A_j}(\L_{\hel}(\A_i)) \\
  \L_{\A_j}(\L_{\hel}(\A_i))^\od &= 0 \\[0.3cm]
  \L_{\A_k}(\L_{\A_j}(\L_{\hel}(\A_i)))^\dia &= 0\\
  \L_{\A_k}(\L_{\A_j}(\L_{\hel}(\A_i)))^\od &= \L_{\A_k}(\L_{\A_j}(\L_{\hel}(\A_i))) ; .
\end{align}

\paragraph{Commutator expressions.}
\begin{align}
  [\K, A] 
  = -\tfrac{\epsi^2}{2}[\Delta, A] 
  = -\tfrac{\epsi }{2} ( \epsi\nabla \cdot A' + A' \cdot \epsi\nabla)
  = -\epsi A'_\nabla 
  = \Or(\epsi).
  \label{eq:commKA}
\end{align}

\vspace{1cm}
\noindent\textbf{Acknowledgment}\\ 
EM thanks financial support of the Swiss National Science Foundation through
a PROMYS Grant (no. IZ11Z0\_166525).


\end{document}